# Nanostructure Phase and interface engineering via controlled Au self-assembly on GaAs(001) surface

*A. Janas[1], B.R. Jany[1*], K. Szajna[1], A. Kryshtal[2], G. Cempura[2], A. Kruk[2], F. Krok[1]*

[1]*The Marian Smoluchowski Institute of Physics, the Jagiellonian University, Lojasiewicza 11, 30-348 Krakow, Poland*

[2]*Faculty of Metals Engineering and Industrial Computer Science, International Centre of Electron Microscopy for Materials Science, AGH University of Science and Technology, 30-059 Krakow, Poland*

Keywords: Gallium Arsenide, AIII-BV, nanostructures, surface diffusion, TEM, SEM

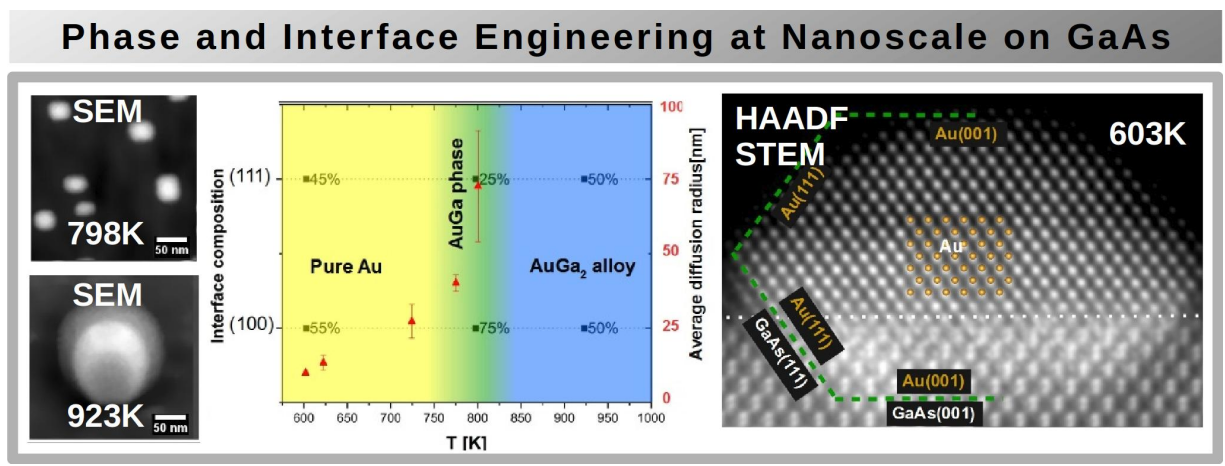

We have investigated the temperature-dependent morphology and composition changes occurring during a controlled self-assembling of thin Au film on the Gallium arsenide (001) surface utilizing electron microscopy at nano and atomic levels. It has been found that the deposition of 2 ML of Au at a substrate temperature lower than 798 K leads to the formation of pure Au nanoislands. For the deposition at a substrate temperature of about 798 K the nanostructures of the stoichiometric AuGa phase were/had been grown. Gold deposition at higher substrate temperatures results in the formation of octagonal nanostructures composed of an $AuGa_2$ alloy. We have proved that the temperature-controlled efficiency of Au-induced etching-like of the GaAs substrate follows in a layer-by-layer manner leading to the enrichment of the substrate surface in gallium. The excess Ga together with Au forms liquid droplets which, while cooling the sample to room temperature, crystallize therein developing crystalline nanostructures of atomically-sharp interfaces with the substrate. The minimal stable cluster of 3 atoms and the activation energy for the surface diffusion $E_d=0.816\pm0.038$ eV was determined. We show that by changing the temperature of the self-assembling process one can control the phase, interface and the size of the nanostructures formed.

---

\*        Corresponding Author e-mail: benedykt.jany@uj.edu.pl

## Introduction

The high carrier mobility and the direct bandgap of Gallium arsenide (GaAs) render it an excellent substrate for the nanoscale electronic industry[1,2]. Especially transistors based on this semiconductor material are the main components of many high-speed and high-frequency electronic devices[3]. Since 1962, when Marshall et al. discovered for the first time stimulated emission in GaAs p-n junction[4]. In recent years for the nanoelectronics industry, semiconductor nanowires are of great interest for the fabrication of new electronic devices. Such systems, based on the vapor-liquid-solid mechanism, are grown by using metal nanoparticles, mostly gold, as a catalyst[5,6]. Therefore, the size, shape, and location of these Au nanoparticles play an essential role in the growth process of semiconductor nanowires. Au is also used in metal/semiconductor systems as an electric contact of the ohmic or Schottky type. For these applications, Au/GaAs systems have been investigated since the 1980s. T. Yoshiie et al.[7] showed that the deposition of a thick film (tens of nm) of Au on a GaAs surface with subsequent annealing leads to the formation of either a rectangular or square-based pyramidal pure Au or alloy $Au_2Ga$ structure. They also observed the appearance of a thin Au-Ga layer only on the structure/substrate interface. It has also been found that the substrate surface preparation plays an essential role in the development of the final Au covered sample morphology. Z. Liliental-Weber et al.[8] already in 1986 had reported that the procedure for GaAs substrate surface preparation (i.e., by ion sputter/annealing in an ultra-high vacuum, by the cleavage of the crystal in UHV or through chemical wet-cleaning) prior to the Au deposition has a strong influence on the physicochemical properties of any formed Au/GaAs interface. The evolution process of the self-assembling of Au on a GaAs surface was also examined by Ming-Yo Li et al.[9] They studied the evolution of a self-assembled Au thin layer on GaAs substrates following annealing in the temperature range of 523 K and 823 K. This research concluded that annealing at about 523 K leads to the formation of Au clusters. Further increase in the annealing temperature leads to the transition from Au clusters (~673 K) to well-formed Au droplet-like nanoislands (~823 K) with a lateral diameter exceeding 100nm. All these studies were performed at sub-micron imaging level for thick Au coverages (~tens of nm) and did not have the chemical sensitivity. Also, they do not provide the information about phase and chemical changes at atomic level.

In this paper, we have investigated the temperature-controlled self-assembly of thin Au layer as deposited on an atomically clean, reconstructed c(8x2) GaAs(001) surface. The morphology of the developed nanostructures was investigated with chemical sensitivity at nanoscale by Scanning Electron Microscopy (SEM) and atomic level by atomically resolved High Angle Annular Dark Field (HAADF) Scanning Transmission Electron Microscopy (STEM). We have found that the substrate temperature plays a significant role in the chemical composition and interface changes at nano and atomic level, which allows for phase and interface engineering. By employing a conventional nucleation theory approach, the activation energy for Au surface diffusion on a reconstructed GaAs(001) surface, as well as the size of the minimal stable cluster, were extracted.

## Experiment / Methods

N-doped epi-ready Gallium arsenide GaAs(001) (MTI Crystal) crystals were cleaned in an ultrasonic bath in isopropanol and mounted on molybdenum plates. Subsequently, the samples were rinsed with ethanol and introduced into a molecular beam epitaxy (MBE) chamber, with a base pressure of $1 \times 10^{-10}$ mbar. The samples were initially outgassed for 2h at 423 K and then cleaned by repeating cycles of ion beam sputtering with 700 eV $Ar^+$ and annealing at 823 K until the clear RHEED (Reflection High Energy Electron Diffraction) pattern of the reconstructed surface was observed. Such a cleaning procedure results in atomically flat surfaces, with a terrace width of 100-200 nm, as already checked with non-contact atomic force microscopy[10]. Au film deposition was performed using molecular beam epitaxy (MBE) from a Knudsen cell kept at 1350 K in the same UHV chamber. The deposition was performed at a rate of 0.1 ML/min (as measured by a quartz microbalance prior to deposition) at different sample temperatures and with a final coverage of 2 ML.

For the examination of developed surface morphology, the samples were transferred, at ambient conditions to the Scanning Electron Microscope (SEM) Quanta 3D FEG. To improve the SEM resolution, the imaging was performed with a decelerated beam voltage of -4kV applied to the sample stage as described in[11]. Furthermore, the surface morphology was also imaged under ambient conditions by means of a 5500 Agilent Atomic Force Microscope in a tapping mode. The size, height and surface density of the formed nanostructures were estimated from SEM and AFM micrographs using free software ImageJ/FIJI[12], Gwyddion[13]. The average diffusion radius of the deposited material was calculated according to $r = \frac{1}{2\sqrt{\rho}}$ where ρ is equal to the average surface density. For the Transmission Electron Microscopy (TEM) investigations thin foils of the samples, covered with a protective layer of thermally evaporated carbon prior to air exposure, were prepared using a focusing ion beam (FIB). Samples so prepared were investigated by an atomically resolved high angle annular dark field scanning transmission electron microscopy HAADF STEM by means of an aberration-corrected probe Titan Cubed G2 60-300 (FEI) microscope operated at 300 kV. The energy dispersive X-ray spectroscopy (EDX) data were obtained using FEI Tecnai Osiris TEM operated at 200 kV electron beam. The collected EDX data were analyzed by the Machine Learning Blind Source Separation method[14] using Non-Negative Matrix Factorization (NMF)[15] as implemented in free software HyperSpy[16] to extract the EDX signal coming from the single nanostructure. Later the composition quantification of the extracted EDX spectra was obtained using the Cliff-Lorimer method[17]. For the clarity only components corresponding to the nanostructures and matrix (i.e. GaAs) were presented in the article. All components for the performed analysis are presented in the Supporting Information.

**Results and discussion**

In Fig.1 a) RHEED pattern of the clean GaAs(001) surface taken along the [110] axis is presented. The main Ga-rich c(8x2) reconstruction spots are clearly visible and residual (n x 6) reconstruction is observed. The evolution of surface morphology after the deposition of 2 ML Au on the reconstructed GaAs(001) substrate surface, maintained at various temperatures is shown in Fig.1 b-f). For those sample temperatures lower than 798 K, nanostructures in the form of nanoclusters are formed (Fig. 1 b-d). They are isotropically distributed on the substrate surface, and their lateral size monotonously increases from 10 nm up to 40 nm, in accordance with the increasing deposition temperature. In addition, the corresponding average height of the nanostructures increases, from 1.5 to 5 nm, as measured with atomic force microscopy. However, the deposition of Au at the substrate temperature T=823 K and higher leads to the formation of nanoislands of an octagonal shape (as shown in Fig.1 e, f). The lateral size of these octagonal-shaped structures is of 100 nm and is almost independent of the increasing substrate temperature. The HR-SEM examination exhibits the facets of the octagonal nanoislands with a top edge oriented along [110] the crystal directions of the GaAs(001) substrate. For the Au deposition performed at 923 K the grown octagonal nanoislands, of an average height around 20 nm, are located on the elongated platforms oriented along [110] the substrate surface direction.

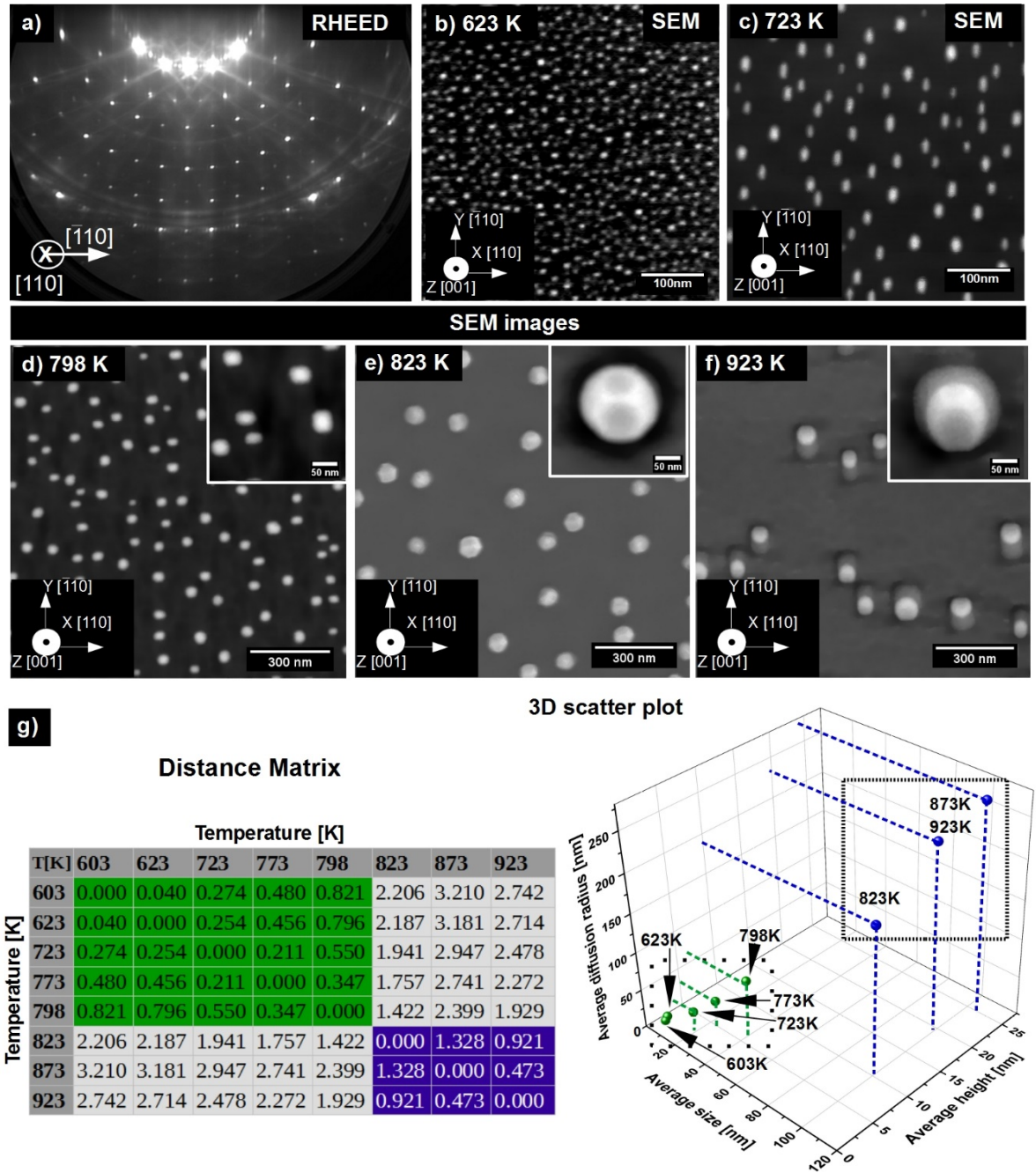

Fig. 1. Surface morphology of a bare GaAs(001) substrate surface and after 2 ML of Au deposition on a GaAs(001) surface at various substrate temperatures. In a) the RHEED diffraction pattern of reconstructed c(8x2) GaAs(001) is shown. In b-f) HR-SEM images of nanostructures formed after Au deposition at the various substrate temperatures are shown (main crystallographic directions are indicated). In g) a three-dimensional plot of the measured nanostructures parameters is shown including the Euclidean distance matrix.

The measured parameters of the formed nanostructures were used as an input to the statistical analysis and to construct a three-dimensional plot containing their average height, size and diffusion radius (Fig. 1g). Each point represents the measured parameters at a certain Au deposition temperature. On the plot, the data are grouped together into two separate groups of points (green and blue clusters), which can be clearly seen, as well as being manifested in the distance matrix where the Euclidean distances between points are calculated in this 3D space. Any small distances between the points indicate that they have very similar properties. The clustering of the data points suggests different nanostructure behavior formed at temperatures below 800 K and above this temperature. To understand these morphological changes, the crystal structure and chemical composition of grown nanostructures were studied.

To investigate in detail the chemical compositions and crystallographic structure of the nanostructures, HAADF STEM together with EDX measurements were performed. In Fig. 2 shown is the transmission electron microscopy measurements from the cross-section of the nanostructure grown after 2 ML Au deposited on GaAs(001) at a temperature of 603 K. The chemical composition of the nanostructure is exhibited by EDX maps (Fig. 2b-d). The enrichment of the nanostructures in Au is shown in Fig. 2b) with a corresponding weak signal of Ga and As. (Fig. 2 c, d). In order to sufficiently distinguish between the EDX signals coming from the nanostructure and the bulk matrix, we used the Machine Learning Blind Source Separation method[14]. This procedure was used to properly separate the signal from the nanostructure and the background (Fig. 2 e-f). Figure 2g) shows the spectrum of the EDX signal from the stoichiometric GaAs substrate acquired in order to check the uncertainty of chemical quantification, where for a perfect quantification the expected result is 50 at.% Ga and 50 at.% As. However, a substrate composition of 52 at.% Ga and 48 at.% As has been obtained, making for uncertainty of ± 2% for our EDX measurements. In analyzing the EDX data, it was found that nanostructures formed at 603 K are made of pure Au. In Fig. 2a) the atomically resolved HAADF STEM image shows that the Au crystalline island is partially buried in the GaAs substrate, this is a consequence of Gold chemical interaction with GaAs substrates[18]. This is also shown by the atomic structural model of Au matching to the experimental HAADF STEM (Fig. 2 a, h)). As indicated in Fig. 2a), the Au nanostructures are in epitaxy with GaAs(001) in the following relationship: (001)Au//(001)GaAs. The principal crystallographic planes of the nanostructure match with the pure Au bulk crystallographic structure ((002) - 2.04Å, (111) - 2.35Å).

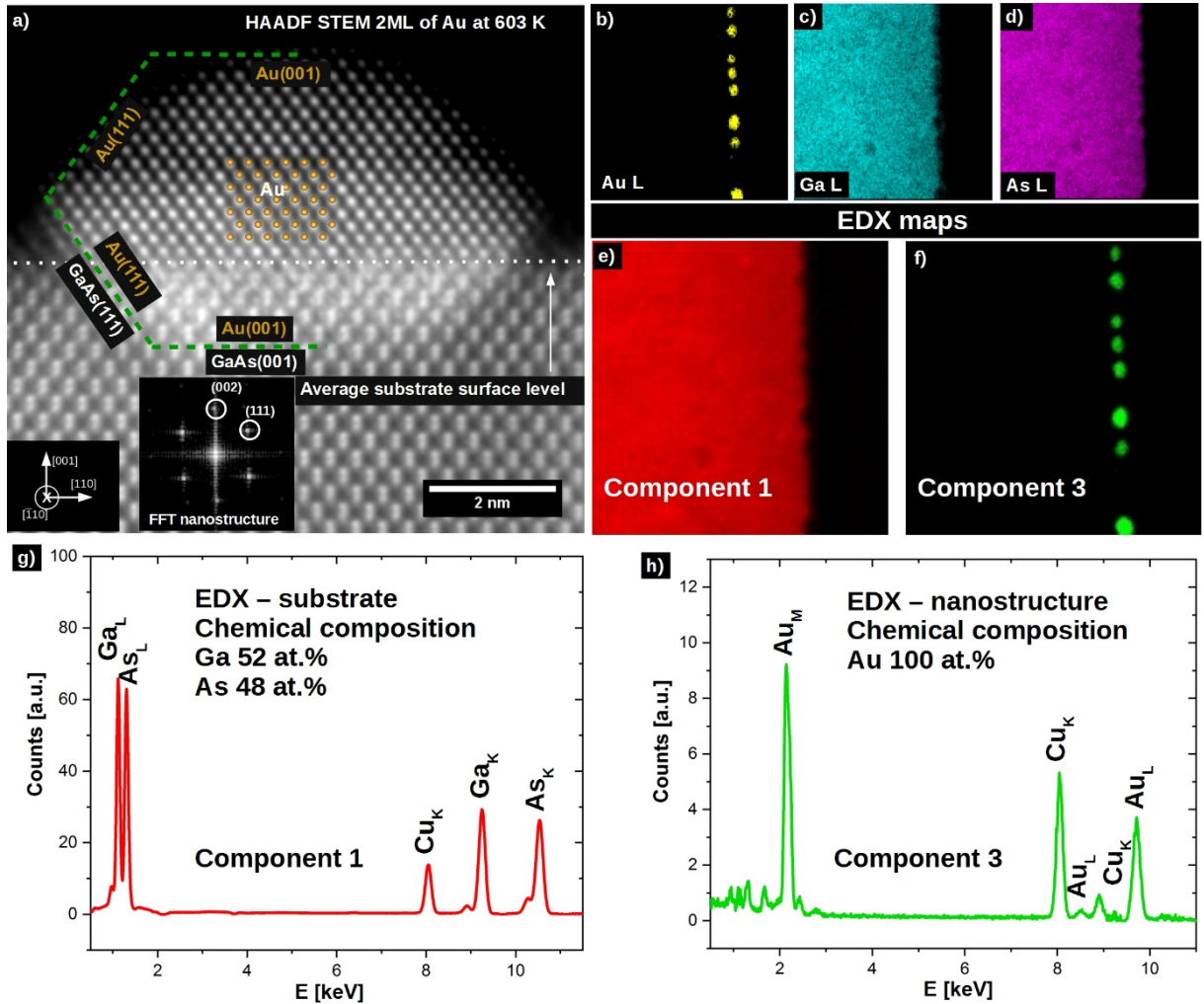

Fig. 2. Nanostructures on GaAs(001) surface resulted from the deposition of 2 ML Au at 600K. In a) an atomically resolved HAADF STEM of nanostructure cross-section with Fourier transform from nanostructure included. The principal crystallographic planes of the substrate and nanostructures are marked. EDX chemical maps of Au, Ga, As are shown in b-d), respectively. Non-negative matrix factorization component maps are shown in e-f) and the corresponding component X-ray spectra in g-h). The origin of the measured $Cu_k$ X-ray signal is the copper holder for the TEM samples.

In Fig.3 presented are HAADF STEM images, together with EDX chemical composition maps of the nanoisland cross section of Au/GaAs(001) grown after Au deposition at 923 K. A well ordered, crystalline structure with atomically sharp interfaces between the grown nanostructure and the substrate is visible. The nanostructure is buried in the GaAs matrix, and as much as 75% of the nanostructure volume is below the substrate surface level. With the detailed analyses of the EDX maps (shown in Fig. 3 b-d), using the same Machine Learning procedure to separate the EDX signal of the nanoisland from the substrate signal (Fig. 2 e-f), we have found that the nanoislands are composed of Au (35(5) at.%) and Ga (65(5) at.%) (Fig 3. i-j). This stoichiometry corresponds to the $AuGa_2$ alloy. Additionally, the Fourier transform analysis taken for the region A (marked in Fig. 3 a), indicates that the interplanar spacing of the nanoisland corresponds to that of the $AuGa_2$ phase: (220) – 2.14Å, (111) – 3.50Å, (002) – 3.03Å. Furthermore, in Fig. 3 g) a HAADF STEM image of the interior part of the nanoisland's cross-section is shown. In this image, it is clearly seen that the bright central columns are surrounded on their side by two "darker" columns. Since the HAADF STEM image contrast is proportional to the atomic number $Z^2$ (thus, providing the chemical sensitivity) the bright columns represent Au, surrounded by two Ga columns. In Fig. 3 h) shown is the appearance of a step-like interface between GaAs(111) and AuGa$_2$(111) as a result of the 7% difference in interplanar spacing.

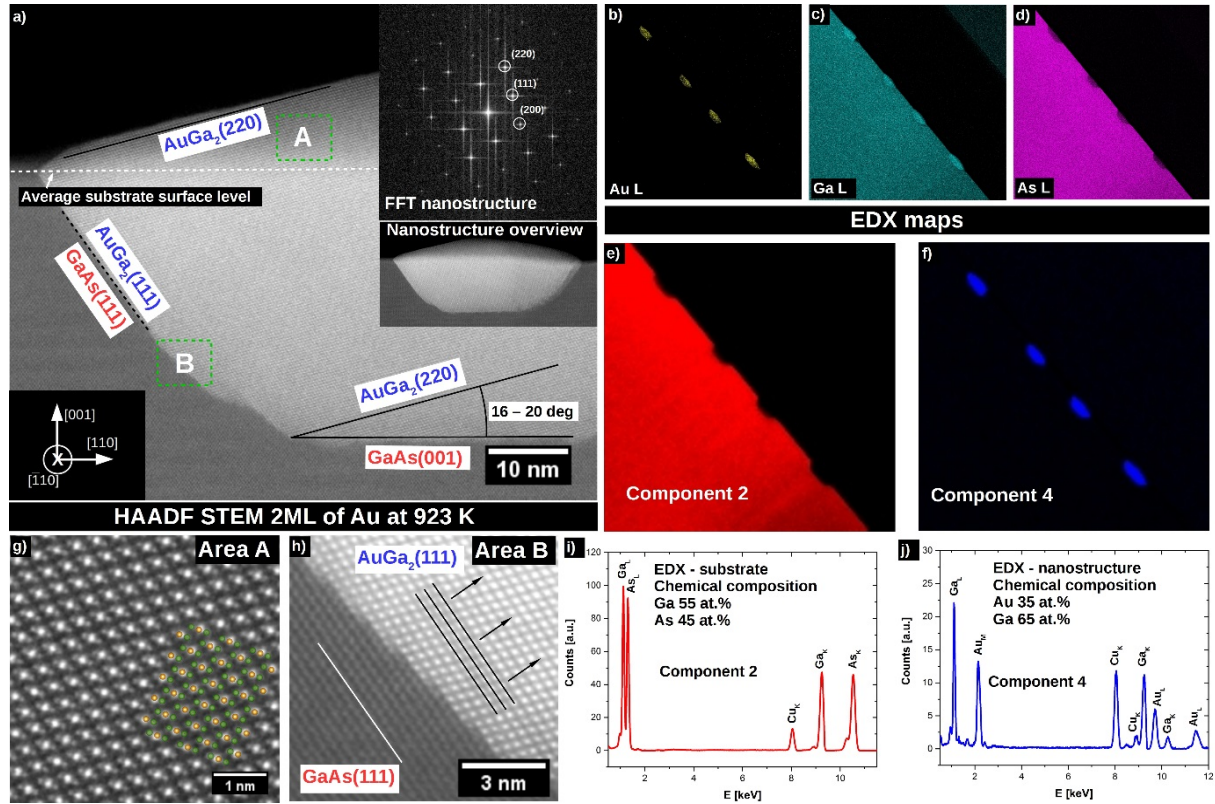

Fig. 3. a) Nanoislands formed after the deposition of 2 ML Au on a GaAs(001) substrate surface at 923 K. The insert in the upper-right side demonstrates a Fourier transform taken from the nano-island. The principal crystallographic planes of the substrate and nanostructure are marked; b)-d) EDX chemical maps of Au, Ga, As, respectively; e)-f) Non-negative matrix component maps and (i-j) corresponding component spectra; g) Detailed HAADF STEM image of the internal nanoisland structure with a fully matched model of AuGa$_2$; h) Characteristic "stepped interface" of GaAs(111) and AuGa$_2$(111) planes, crystallization direction is indicated by black arrows.

The origin of the metallic nanostructure buried into GaAs matrix with atomically sharp interfaces can be interpreted as follows: during the deposition of Au atoms on the GaAs substrate the bonds of the matrix are broken and an etching process similar to the metal-assisted etching process[19] occurs. According to the Au-Ga phase diagram[20], at these elevated temperatures, alloy liquid droplets are formed by the interaction of Au with released Ga atoms. Since no As content is detected in the nanostructures grown, we assume that it is released into the vacuum as a gas phase. Then during cooling, the sample crystallization process of the liquid droplets begins. Due to the lowest energy of the (111) crystallographic planes[21] the system recrystallizes by an exhibiting of (111) planes in the direction marked by the black arrows in Fig. 3h). The formed AuGa$_2$ nanoislands are in epitaxy with GaAs(001) in the following relationship: (110)AuGa$_2$ tilted by 18°//(001)GaAs. Also, the long-range crystallographic ordering of the AuGa$_2$ nanoislands (as proved in Fig. 3 g,h) indicates that liquid droplets are formed when the Au film is deposited and then they recrystallize when the sample is cooled down to RT.

The scenario presented above was directly verified by a dedicated experiment in which we performed in situ RHEED measurements in order to gain direct insight into the dynamics of the nanostructure formation process. Fig. 4 illustrates the results of the examination of nanoislands formation after the deposition of 2 ML Au on a reconstructed GaAs(001) substrate surface at 798 K. In Fig.4a) shown is a HAADF STEM image of the nanoislands cross-section, together with EDX chemical composition maps. About 40% of the nanostructures volume is located below the average substrate surface level. The Machine Learning analysis of EDX signals provides that the nanostructure are an AuGa alloy composed of Au and Ga at a 1:1 ratio. Fig. 4c) depicts the evolution of the RHEED pattern during the formation of the metallic nanostructures on the reconstructed GaAs(001) surface. After the Au deposition with the subsequent cooling to RT, new diffraction spots in the RHEED pattern appear (some of them indicated with white arrows). These spots are not sensitive to the sample rotation in the XY plane, which indicates that this is a transmission pattern from 3D objects[22]. The graph of the temperature dependent intensity of one of these diffraction spots (indicated by a red dotted circle in Fig. 4c) shows the evolution of the system under investigation when cooling down the sample. At a temperature of about 600 K, the signal significantly increases which we correlate with the nanostructures crystallization (see the movie in Supplementary Materials). Thus, determined was the threshold temperature of AuGa nanoisland crystallization equal to 596±10 K. A Similar temperature of such a phase transition for the system of a 1 nm thick layer of Au deposited at RT on a GaAs substrate was measured by J.C. Harmanda et al.[23]

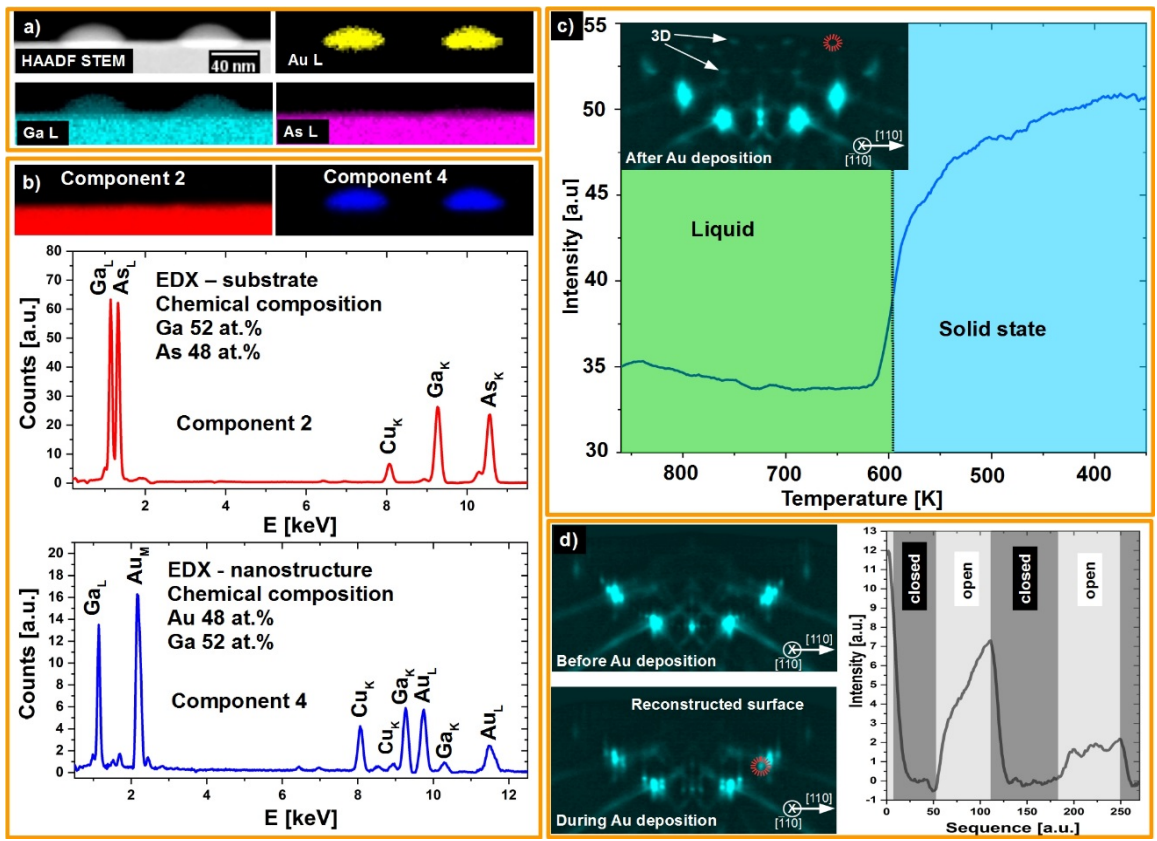

Fig. 4. Results for nanostructures formed after the deposition of 2 ML Au on a GaAs(001) substrate surface at 798K; a) HAADF STEM image of the nanostructure together with the EDX chemical maps of Au, Ga, As, respectively; b) Non-negative matrix factorization component maps and their corresponding component spectra; c) RHEED diffraction patterns after Au deposition on a reconstructed GaAs(001) surface with marked transmission spots corresponding to 3-dimension nanostructures. The intensity of one of the transmission spots, selected by a red dashed circle, was used to estimate the dynamic of nanostructure crystallization. d) RHEED diffraction pattern before and during Au deposition on a reconstructed GaAs(001) surface. The signal of one of the diffraction spots, marked by a red dashed circle, was used to analyze signal intensity as a function of Au deposition/no-deposition.

Additionally, in the RHEED experiment, we also observed changes in the diffraction pattern of the reconstructed substrate surface during the Au deposition. Fig. 4d) shows RHEED patterns before and during Au deposition on a reconstructed GaAs(001) surface at an elevated temperature. Before Au evaporation, the (nx6) reconstruction of the substrate surface is clearly seen. While the Au deposition, at a coverage of 0.1 Au ML the (nx6) reconstruction spots disappear and with a coverage of 0.25 Au ML a new substrate surface (nx4) reconstruction appears with the corresponding RHEED diffraction spots as indicated by a red dashed circle in Fig. 4d). This Au-induced reconstruction is metastable and is present only if the Au is supplied. When the shutter of the effusion cell is closed, new RHEED diffraction spots disappear. This behavior is clearly demonstrated in the graph of intensity of one of the new diffraction spot versus periodic closing/opening of the Au source. Closing the Au source causes the diffraction spot to disappear with a time constant of 10 sec. This new RHEED pattern appears again at the reopening of the Au source, although with a lower intensity probably due to the increasing participation in the evolving morphology of the developing nanoislands. Analysis of the RHEED pattern proves that the metastable, Au-induced reconstruction is of a characteristic distance of 1.6 nm along [110] the GaAs crystallographic direction. We observe this new RHEED diffraction pattern only in a narrow temperature range close to 798K. Having in mind that during the process of Au deposition, the AuGa droplets grow (expand), we can conclude that the simultaneous coexistence of the reconstruction of the substrate surface indicates a homogeneous, preferential etching of the substrate in a "layer by layer" manner.

The results of the morphology and chemical composition examination of the 2 ML Au/GaAs(001) system in the process of thermally-controlled Au self-assembly are shown in Fig. 5a). Two distinct temperature regimes, separated at T~810 K and associated with the nanoislands shape change, are visible: the first regime with a linear increase in lateral nanostructures size and the second regime, at which the size remains constant at ~100 nm. Nanostructures formed during the first temperature regime exhibit a monotonous increase in their lateral size, from 10nm at T=603 K (pure Au phase) to 40 nm at T=798 K, which is already AuGa alloy phase. Furthermore, in this temperature regime, we observe a linear trend in the average diffusion radius in the Arrhenius plot (Fig. 5c), corresponding to the case of morphology developing as a result of pure surface diffusion processes. The first phase transformation is observed at a temperature of around 798 K, for which the AuGa alloy appears. In Fig. 5d) it is clearly seen that for AuGa alloy nanostructures the (001) interface contribution becomes larger than in cases where the nanostructures were of pure Au. The contribution increases from 55% for pure Au to 75 % for the AuGa phase as illustrated in Fig 5b). On the other hand, for the second temperature regime, enhanced effective breaking of GaAs bonds by Au atoms occurs, because a large part of the formed nanostructures is buried under the average substrate surface level. The nanostructures are formed of an $AuGa_2$ alloy with their lateral size being ~100 nm. The $AuGa_2$ alloy formation was confirmed by SEM EDX chemical composition analysis supported by Machine Learning[24]. Moreover, the dependence of the average diffusion radius differs from the linear trend in the Arrhenius graph (Fig. 5c). The contribution of individual crystallographic planes to the whole interface between the nanostructure and the substrate also changes, in this case, it consists of about 50% of the (111) plane and 50% of the (001) plane. This allows for a natural interface size and phase type control through selecting an appropriate temperature regime for the self-assembly process. Thus, taking into account that the dominant part of $AuGa_2$ nanostructures (Fig. 5d) is built into the matrix, but also that the Arrhenius dependence is not preserved, it is evident that in this temperature regime a competitive process to the surface diffusion process takes place. We identify this with a more effective breaking of the bonds of matrix atoms by Au atoms, which results in an intensive etching of the substrate in the position of an initial stable cluster: leading eventually to the formation of liquid droplets.

For the case of nanostructures grown in the first temperature regime, with the help of Arrhenius dependence (Fig. 5c) we are able to calculate the activation energy for the surface diffusion process. In the conventional nucleation theory approach with an increasing average surface diffusion radius, the density of the nanoislands formed should decrease if no-reaction between the components of the considered system is assumed[25,26]. We have used this theory to calculate the critical nucleus size and activation energy, regardless of our active chemical system as this has already been successfully shown for a chemically interacting Fe/Si system[27]. According to this nucleation theory the density of the formed structures ρ decrease exponentially with increasing temperature and is described by the following equation:

$$\rho = \eta \left(\frac{F}{D_0}\right)^{\frac{i}{i+2.5}} \exp\left(\frac{E^*}{k_B T}\right),$$

Where $\eta = 0.06$ and represents a dimensionless constant comprising the coverage dependence[25], F=0.002 ML/s depicts the deposition flux, and $D_0$ is the surface diffusion factor. Due to the motion on the atomic scale, surface areas are considered as unit cells, $D_0$ becomes the number of available unit cells visited by the adatom per unit time[26]. In our case surface diffusion factor is described by equation:

$$D_0 = \frac{a^2 v_0}{z},$$

where "a" is jumping distance and is of the order 1 nm, hence order 2.5 lattice distances (lattice distance corresponds to distance between atoms on GaAs(111) surface – 0.399 nm), $v_0$ is the vibration frequency of the atom – $10^{12}$/s. Term "z" describes the number of neighboring adsorption sites (4 for square unit cells). Then, surface diffusion factor is $1.57 \cdot 10^{12}$/s. The largest unstable cluster, the "i" term, indicates the number of atoms for which the addition of one extra atom starts to be stable. E* is the effective diffusion barrier, which is composed of two terms: $E_d$ – activation energy for the diffusion and $E_b$ – critical cluster binding energy, providing E*=(i$E_d$ + $E_b$)/(i+2.5), where 2.5 is scaling constant for 3D islands. The total effective diffusion barrier E* is calculated from a linear dependence in the Arrhenius plot - ln(ρ) as a function of $1/k_b T$ (Fig. 5c) where Coefficient of Determination ($R^2$ from the linear fit) is 0.9933 and Chi-Sqr/NDF is equal to 1.017. Critical cluster size can be evaluated from the Arrhenius plot fit extrapolation at 1/T approaching to 0. From this approach, it has been found that E*=0.633±0,029eV and the critical cluster size of i = 2.429±0.075 i.e. the cluster consist of 3 atoms is stable. Furthermore, taking the binding energy of a critical $Au_2$ cluster ($E_{Au2}$= 1.22 eV[28]) and i=2, which represents number of atoms in critical cluster, obtained has been the activation energy for the Au diffusion on the GaAs(001) c(8x2) surface equal to $E_d$= 0.816±0.038 eV which is in good agreement with 0.8 eV obtained by T. Yoshiie and cowerkers[7]. The estimated uncertainty of the derived activation energy and critical cluster size reflects only the variability and the uncertainty of our experimental data.

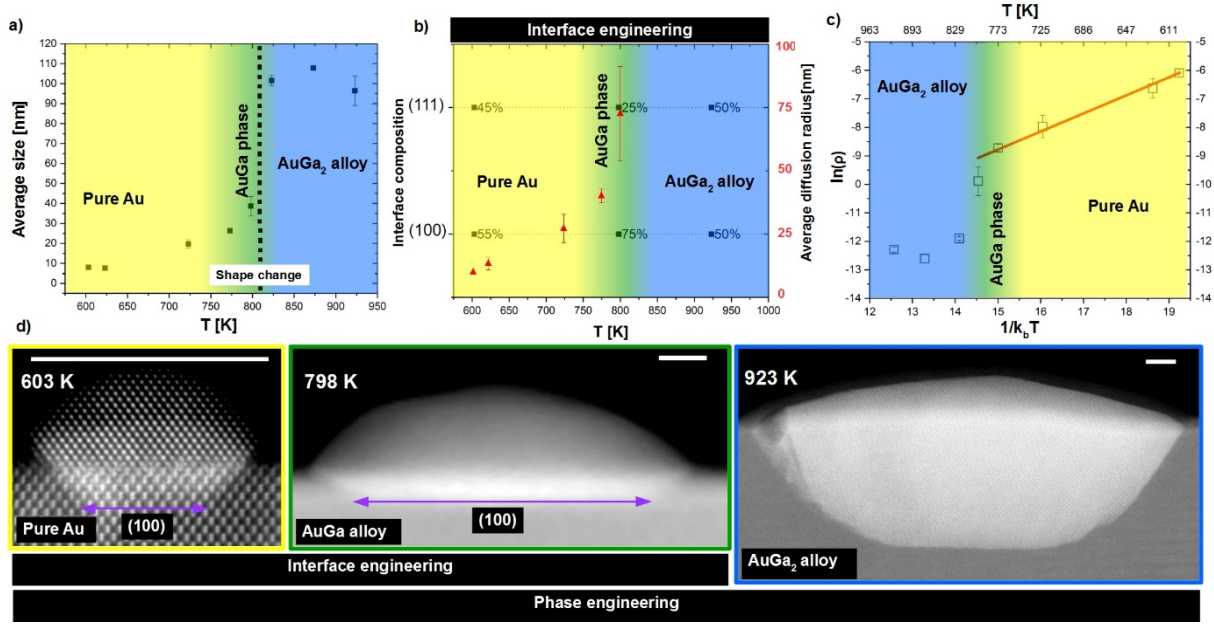

Fig. 5. Interface and phase engineering for the Au/GaAs (001) system. a) Average nanostructure size as a function of the GaAs(001) substrate temperature during Au deposition, the temperature regimes of occurrence of different phases are indicated. b) Nanoisland/GaAs substrate interface composition. c) Arrhenius plot of the nanostructures surface density. d) HAADF STEM images represent the evolution of the interface formation, scale bars correspond to 8nm, interfaces marked by purple arrows.

**Conclusions**

We have systematically investigated the temperature-controlled dynamics of the morphology and composition changes of metallic nanostructures formed upon the deposition of a thin, 2ML layer of Au on a GaAs(001) surface. It has been found that for the deposition temperatures below 823 K Au nanoislands grow with a size and surface density which depend on the substrate temperature. In the narrow substrate temperature window, around 798 K, we have observed the formation of nanostructures of the AuGa phase; whereas for higher deposition temperatures, well-shaped octagonal nanoislands of 100 nm in size, composed of the $AuGa_2$ phase, developed. The composition of these nanostructures and the fact that they are "buried" into the substrate, point to an Au-induced enhance bond breaking of the substrate atoms and the formation of Au-Ga alloy liquid droplets during the deposition process. We have found that during the cooling of the sample at a temperature of 798 K the droplets crystallize leading to the formation of metallic nanostructures of a crystallographic order as well as atomically sharp nanostructure/substrate interfaces. Applying conventional nucleation theory, the activation energy for surface diffusion on a GaAs(001) surface of 0.816±0.038 eV and the minimal stable cluster consisting of three Au atoms has been obtained. We have shown that by changing the temperature of the self-assembling process, one could control the nanostructure interface size and the nanostructure phase, which could be used for the engineering of desired nanoelectronic devices based on GaAs.

**Acknowledgments**

We gratefully acknowledge the help of Prof. R. Abdank Kozubski in fruitful discussions. Support by the Polish National Science Center (UMO- 2018/29/B/ST5/01406) and from the Polish Ministry of Science and Higher Education under the grant 7150/E-338/M/2016 is also acknowledged.

## Competing financial interests

The authors declare no competing financial interests.